\documentclass[10pt, conference, compsocconf]{IEEEtran}

\usepackage[cmex10]{amsmath}
\usepackage{amsfonts}
\usepackage[tight,footnotesize]{subfigure}
\usepackage{chngcntr}
\counterwithout{subsubsection}{subsection}

\usepackage{algorithm}
\usepackage[noend]{algorithmic}
\usepackage{booktabs}
\usepackage{tabularx}
\usepackage{hyperref}

\usepackage{graphicx}

\newenvironment{Algorithm}[1][tbh]%
  {\centering
  \begin{minipage}{#1}
  \begin{algorithm}[H]}%
  {\end{algorithm}
  \end{minipage}\par
  \vspace{\belowdisplayskip}}

\newcommand{\spark}[1]{\texttt{\small #1}}

\begin{document}
%
\title{Scalable Manifold Learning for Big Data with Apache Spark}


\author{\IEEEauthorblockN{Frank Schoeneman}
\IEEEauthorblockA{Department of Computer Science \& Engineering\\
University at Buffalo\\
Buffalo, New York, USA\\
fvschoen@buffalo.edu}
\and
\IEEEauthorblockN{Jaroslaw Zola}
\IEEEauthorblockA{Department of Computer Science \& Engineering\\
Department of Biomedical Informatics\\
University at Buffalo\\
Buffalo, New York, USA\\
jzola@buffalo.edu}
}

\maketitle

\begin{abstract}
Non-linear spectral dimensionality reduction methods, such as Isomap, remain important technique for learning manifolds. However, due to computational complexity, exact manifold learning using Isomap is currently impossible from large-scale data. In this paper, we propose a distributed memory framework implementing end-to-end exact Isomap under Apache Spark model. We show how each critical step of the Isomap algorithm can be efficiently realized using basic Spark model, without the need to provision data in the secondary storage. We show how the entire method can be implemented using PySpark, offloading compute intensive linear algebra routines to BLAS. Through experimental results, we demonstrate excellent scalability of our method, and we show that it can process datasets orders of magnitude larger than what is currently possible, using a 25-node parallel~cluster. 
\end{abstract}


%
\IEEEpeerreviewmaketitle

\section{Introduction}

The vast majority of the current big data, coming from, for example, high performance high fidelity numerical simulations, high resolution scientific instruments (microscopes, DNA sequencers, etc.) or Internet of Things streams and feeds, is a result of complex non-linear processes. While these non-linear processes can be characterized by low dimensional manifolds, the actual observable data they generate is high-dimensional. This high-dimensional data is inherently difficult to explore and analyze, owing to the {\em curse of dimensionality} and {\em empty space phenomenon} that render many statistical and machine learning techniques (e.g. clustering, classification, model fitting, etc.) inadequate. In this context, non-linear spectral dimensionality reduction has proved to be an indispensable tool~\cite{Lee2007}. The non-linear spectral dimensionality reduction methods rely on a spectral decomposition of a feature matrix that captures properties of the underlying manifold, and effectively bring the original data into a more human-intuitive low dimensional space that makes quantitative and qualitative analysis of non-linear processes possible (e.g., by enabling visualization). However, the leading manifold learning methods, such as considered here Isomap~\cite{Tenenbaum2000}, remain infeasible when presented with the current large-scale data. This is because of both computational and memory complexity that are at least quadratic in the number of input points.

In this paper, we propose a distributed memory framework to address complexity of the end-to-end exact Isomap under Apache Spark model. We show how each critical step of the Isomap algorithm can be efficiently expressed under the basic Spark model, without the need to provision data in the secondary storage. To achieve efficient and portable software realization, we implement the entire method using PySpark, and we offload compute intensive linear algebra routines to high performance BLAS library. Through experimental results, we demonstrate that the resulting software exhibits excellent parallel scalability, and even on a small parallel cluster can be used to process datasets orders of magnitude larger than what is possible withe the current methods, without sacrificing quality. Our specific contributions include Apache Spark-based efficient direct $k$NN solver based on 1D data decomposition, scalable all-pairs shortest-path (APSP) solver leveraging ideas from communication avoiding algorithms, and practical eigendecomposition solver based on power iteration. All components are combined into end-to-end workflow realizing in-memory Isomap.

The reminder of the paper is organized as follows. In Section~\ref{sec:prelim}, we formally state the problem and briefly introduce the basic Isomap algorithm. In Section~\ref{sec:sparkiso}, we provide detailed description of our proposed approach, focusing on each individual Isomap component and its realization in Apache Spark. In Section~\ref{sec:results}, we showcase the resulting software on several benchmarks, assessing accuracy and parallel scalability, and we demonstrate its practical value by analyzing over 50000 EMNIST images. Finally, in Section~\ref{sec:relwork}, we survey related work, and conclude the paper in Section~\ref{sec:end}. 

\section{Preliminaries}\label{sec:prelim}

In manifold learning, we assume that a given dataset $X = \{x_0, x_1, ..., x_{n-1}\}$ of $n$ points, where each $x_i \in \mathbb{R}^{D}$, is sampled from some manifold $\mathcal{M}$ of latent dimension~$d$, $d \ll D$. Our goal is to find a set $Y = \{y_0, y_1, ..., y_{n-1}\}$ such that $y_i\in\mathbb{R}^{d}$ and $\forall_{i, j}$ $d_{\mathcal{M}}(x_i, x_j) \approx | y_i - y_j |_{h}$, where $d_{\mathcal{M}}(\cdot)$ measures the actual distance between pair of points on $\mathcal{M}$, and the relation $|a-b|_h$ captures some property between points $a$ and $b$. In this work, we focus on the case where $|\cdot|_h$ is $\lVert \cdot \rVert_{2}$, that is, we seek a $d$-dimensional representation,~$Y$, of~$X$ that preserves manifold distances.

The state-of-the-art method for geodesic distance preserving manifold learning is Isomap~\cite{Tenenbaum2000}. The method falls under the category of non-linear spectral dimensionality reduction techniques~\cite{Lee2007}, and it requires a series of data transformations shown in Alg.~\ref{alg:isomap}. In the first step, the dataset $X$ is used to construct a neighborhood graph, $G$, where each node of the graph is connected to its $k$ nearest neighbors ($k$NN), with respect to the Euclidean metric. Here $k$NN distances capture local manifold distances in the neighborhood of each data point. This step requires $\mathcal{O}(n^{2})$ distance computations, thus having complexity $\mathcal{O}(Dn^{2})$. In the second step, the neighborhood graph $G$ is analyzed to construct a feature matrix, $A$, that captures the information to be preserved in dimensionality reduction. Because shortest paths induced by neighbor relations have been shown to be a good approximation of the actual manifold distance between points~\cite{Bernstein2000}, $A$ stores the shortest paths between all pairs of points in the neighborhood graph $G$. Consequently the cost of this step is $\mathcal{O}(n^{3})$. The penultimate step is a spectral decomposition of $A$. The resulting top $d$ eigenvectors, $Q_d$, and eigenvalues, $\Delta_d$, are required as their product yields the final output set $Y$. The spectral decomposition step has complexity $\mathcal{O}(n^{3})$. We note that before spectral decomposition, the feature matrix $A$ is usually double-centered to ensure that the origin is contained in the final output (instead of being an~affine~translation).

\begin{Algorithm}[\columnwidth]
    \caption{\textsc{Isomap}}
	\begin{algorithmic}[1]
        \REQUIRE $X$, $d$, $k$
        \ENSURE $Y$
        \STATE $G$ = \textsc{kNN}($X$, $k$)
        \STATE $A$ = \textsc{AllPairsShortestPaths}($G$)
        \STATE $D$ = \textsc{DoubleCenter}($A^{\circ 2}$)
        \STATE $(Q_{d},\Delta_{d})$ = \textsc{EigenDecomposition}($D$)
        \STATE $Y$ = $Q_d \cdot \Delta_d^{\circ \frac{1}{2}}$
        \RETURN{$Y$}
    \end{algorithmic}
    \label{alg:isomap}
\end{Algorithm}
 
While Isomap is the method of choice in many practical applications~\cite{Lee2007}, it is too computationally and memory intensive for even modest size datasets. For example, commonly available sequential implementations in Matlab and Python scale to datasets with $n=4000$ points. In contrast, datasets emerging in scientific applications, e.g.~\cite{Turnbaugh2007, Marchand2009,Thirion2004, Li2012}, routinely involve hundreds of thousands of points. 

\section{Proposed Approach}\label{sec:sparkiso}

The computational complexity of Isomap arises from the cost of both construction and spectral decomposition of the feature matrix. The memory cost comes from the $\Theta$($n^2$) size of the feature matrix that has to be maintained throughout the process. To alleviate these complexities, several approximate methods have been proposed~\cite{deSilva2003,Talwalkar2008}. However, these approaches do not provide exactness guarantees, or are tailored for large-scale HPC systems~\cite{Samudrala2015}.

In our proposed approach, we focus on exact solution targeting Apache Spark clusters that are easy to deploy, and are accessible to many potential~end-users. Our main idea is to cast the Isomap workflow into the Apache Spark model, such that the intermediate data is never explicitly provisioned to the persistent storage, and each step of the workflow is efficiently expressed via Spark transformations implemented in PySpark and Python, with computationally intensive algebraic routines offloaded to a dedicated BLAS engine (e.g., Intel MKL).

\subsection{\texorpdfstring{$k$NN}{kNN} Search}\label{ssec:knn}

The first step of Isomap is $k$ nearest neighbors search. The direct approach to $k$NN is for each point to compute the distance to each of the $n-1$ others, recording the $k$ minimum in the process. This requires $\Theta$($n^2$) comparisons and $\Theta$($nk$) space. Theoretically, this approach can be improved by the use of spatial data structures such as $k$-$d$ trees, quad-trees, R-trees, or Voronoi diagrams~\cite{Bentley1975,Kim2016}. However, due to the \emph{curse of dimensionality}, the performance of these data structures quickly deteriorates to the direct method as dimensionality $D$ increases~\cite{Weber1998}. Therefore, in our approach we propose scalable Spark realization of the direct~$k$NN~method.

\begin{figure}
  \centering
  \includegraphics[scale=0.875]{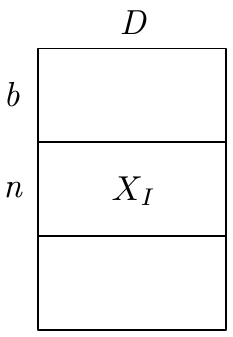}~~
  \includegraphics[scale=0.875]{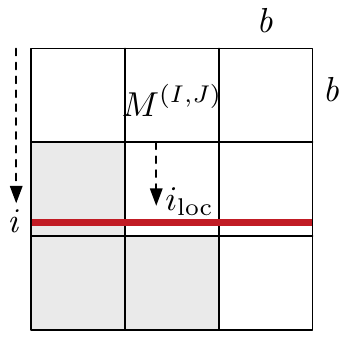}~~
  \includegraphics[scale=0.875]{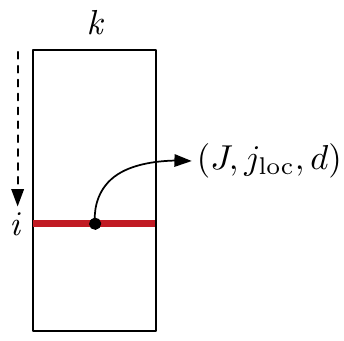}
  \caption{Left: 1D decomposition of the input data $X$. Middle: Matrix $M$ and example row $i$. Right: $k$NN RDD and element of list $L_k$.}\label{fig:KNN}
\end{figure}

We first 1D-decompose the input data $X$ into $\lceil \frac{n}{b}\rceil$ logical blocks (see Fig.~\ref{fig:KNN}). We achieve this by loading the entire $X$ into RDD, such that single point is stored as one 1D NumPy array, and then executing \spark{combineByKey} transformation. The transformation assigns to a point its block identifier, and points are grouped into a 2D NumPy array corresponding to a block. Hence, at the end of this step, each point in $X$ is assigned to a unique block $X_I$, for $0 \leq I < q = \lceil \frac{n}{b} \rceil$. By decomposing $X$ in this way, we can exploit symmetry of the distance matrix over $X$, and efficiently compute only its upper-triangular portion. This is contrary to the standard and expensive approach based on Spark \spark{cartesian}, which produces RDD elements $(I, J)$ and $(J, I)$, requiring additional \spark{filter} operation to obtain the final upper-triangular form. Using \spark{flatMap} transformation, we enumerate all block pairs $(X_I, X_J)$, where each $X_I$ is paired with all $X_J$, for $J \geq I$. The resulting RDD, stores each block pair as a tuple $((I,J), (X_I,X_J))$. Although this induces data replication, it also directly exposes parallelism in computing the distance matrix that dominates the computational work of $k$NN search. We exploit this parallelism when materializing the upper-triangular distance matrix, $M$. Specifically, we apply \spark{map} transformation that for each pair $(X_I, X_J)$ carries all-pairs distance computations to yield sub-matrix $M^{(I, J)}_{i,j} = \lVert x_{i} - x_{j} \rVert_{2}$, $\forall x_{i}\in X_I, x_{j}\in X_J$. Here, computing $\lVert x_{i} - x_{j} \rVert_{2}$ is delegated to SciPy, which offloads distance computations to the efficient BLAS library (e.g., Intel MKL). At the end, the entire block $M^{(I, J)}_{i, j}$, stored as 2D NumPy array, becomes a single element of the resulting RDD with $M$. The RDD is persisted in the main memory for further~processing.

The next step is to identify nearest neighbors of each $x_i \in X$ with respect to matrix $M$. When $M$ is in the block form, each row $i$ of $M$ is scattered across multiple column-blocks~$J$ (see Fig.~\ref{fig:KNN}). With each such chunk of row $i$ we can associate the same pair $(I, i_{\textrm{loc}})$, where $I$ identifies row-block containing $i$, and $i_{\textrm{loc}}$ is local identifier of row $i$ within row-block $I$ (since within a block, rows are indexed from~0, and we have that $i = I\cdot b + i_{\textrm{loc}}$). We exploit this property, and for each point $x_i$ we identify the minimum $k$ elements as follows. First, within each block $M^{(I,J)}$ for every $i_{\textrm{loc}}$ we identify a list of $k$ minimal values, $L_k$. We achieve this in parallel via \spark{flatMap} transformation that employs standard heap-based algorithm to maintain minimal values, and yields tuples $((I, i_{\textrm{loc}}), L_k)$. Here, $L_k$ stores coordinates of the selected minimal values, and the values themselves. Recall that we do not explicitly represent $M^{(I, J)}$ for $I > J$. Therefore, to obtain row minima for blocks under diagonal of $M$, our \spark{flatMap} routine considers also transpositions $(M^{(J,I)})^{T}$. Next, using \spark{combineByKey} transformation, we reduce all local minima, contained in lists $L_k$, into the global $k$NN list for each~$x_{i}$ (see Fig.~\ref{fig:KNN}).

While at this point we could finalize $k$NN search, we apply one more round of transformations, to efficiently convert $k$NN into the neighborhood graph $G$, such that it is ready for processing via all-pairs shortest-paths solver. The idea is to reuse data blocks maintained by RDD storing~$M$, to store~$G$. To this end, we apply \spark{map} transformation to the $k$NN and for each~$x_i$ we produce key-value pair $((I, J), (i_{\textrm{loc}}, j_{\textrm{loc}}, d))$, where tuple $(J, j_{\textrm{loc}}, d)$ identifies nearest neighbor of $x_i$. This enables us to associate each $k$NN distance with the sub-block of $M$ to which it belongs. Thus, we use \spark{union} to combine the resulting RDD with the original block matrices of $M$, and we follow with \spark{combineByKey} transformation. The transformation fills blocks $(I,J)$ of $M$ with $\infty$, and then sets the actual neighbor distances according to $(i_{\textrm{loc}}, j_{\textrm{loc}}, d)$. The resulting neighborhood graph, $G$, is stored as RDD of blocks of size $b \times b$, in exactly the same way as $M$, benefiting from the fact that $M$ is persistent and minimizing NumPy arrays reallocation.

\begin{figure}[t]
   \centering
   \includegraphics[scale=0.9]{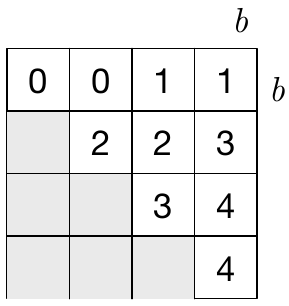}
   \caption{Example of assignment of logical blocks to RDD partitions. Number inside a block indicates RDD partition storing the block.}
   \label{fig:rddblock}
\end{figure}

The 1D decomposition we use in our $k$NN solver, and the induced 2D decomposition of matrix $M$, are logical and separate from the physical RDD partitioning employed by the Spark runtime. Moreover, instead of relying on the default Spark partitioner, we use a custom partitioner tailored for managing upper-triangular matrices. This is motivated by the performance considerations. The partitioner assigns an integer to each block of the matrix in the row-major fashion. Hence, when the number of RDD partitions $p'$ is equal to the number of logical blocks, the assignment is exactly the upper-triangular row-major index for block $(I, J)$. When this number is less than $\frac{q\cdot (q+1)}{2}$, we store $B = \frac{Q}{p'}$ blocks per partition, where $Q = \frac{q\cdot (q+1)}{2}$. Specifically, we assign the first $B$ blocks to partition 0, the next $B$ blocks to partition 1, and so on (see Fig.~\ref{fig:rddblock}). In our experiments, we found that this strategy of assigning multiple blocks to single RDD partition provides better data locality than the default partitioner or \spark{GridPartitioner} found in the MLlib library~\cite{Meng2016} (a popular machine learning library for Apache Spark). This is because neighboring blocks are stored near one another in RDD, which leads to improved data locality and reduced data shuffling. Therefore, we employ the same custom partitioner in the APSP solver (see below).

\subsection{All-pairs Shortest-paths Computation}\label{sssec:apsp}

Given the neighborhood graph $G$, the next step in the Isomap workflow is to solve all-pairs shortest-paths (APSP) in $G$. Recall that shortest paths approximate manifold distances between data points~\cite{Bernstein2000}, and hence they will form the feature matrix $A$. 

Two commonly used algorithms for APSP are Dijkstra and Floyd-Warshall~\cite{Cormen2009}. However, both methods are ill-suited for Spark model, owing to low computation rate compared to the data movement rate. An alternative approach involves a special case of the matrix power $A^{n}$~\cite{Kepner2011}. Consider matrix multiplication over the semiring $\overline{\mathbb{R}}_{+} = \mathbb{R}_{+} \cup \{+\infty\}$ with additive and multiplicative operations, $\oplus$ and $\otimes$. The additive operation is \emph{minimum} and the multiplicative operation is \emph{scalar addition}, that is $x\oplus y \mapsto \min\{x, y\}$ and $x\otimes y \mapsto x + y$. In this way we reduce APSP to repeated matrix multiplication of $G$ by itself (we refer to this method as \emph{min-plus}). This is still not ideal: regular data movement patterns like those in matrix multiplication are unsuitable for data shuffles that occur under the Spark model. In fact, matrix multiplication is yet to see efficient realization in Spark~\cite{Bosagh2016}. Therefore, we take a middle-ground approach.

We cast an iterative communication-avoiding APSP algorithm by Solomonik et al.~\cite{Solomonik2013}, which is based on the block formulation of Floyd-Warshall~\cite{Venkataraman2003}, into Spark model, and we use min-plus to batch update $b\times b$ blocks of shortest paths values. It turns out that communication-avoiding algorithms, technique well known in High Performance Computing, enable us to minimize data shuffles, leading to efficient Spark realization. We note that the correctness of this algorithm follows directly from that of computing the transitive closure, and has been discussed, e.g., in~\cite{Solomonik2013,Venkataraman2003}.

\begin{figure}[t]
   \centering
   \includegraphics[scale=0.9]{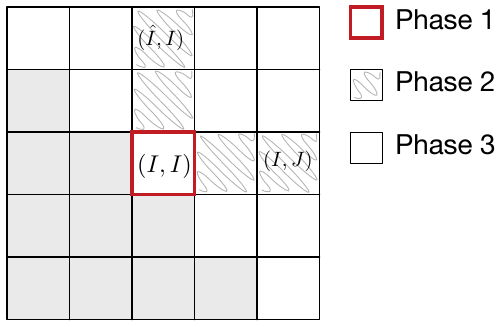}
   \caption{Phases of computation for iteration $\hat{I}$ along critical path defined by the block diagonal. In Phase~1, sequential Floyd-Warshall is performed on diagonal block $(I,I)$. In Phase~2, the solution of the diagonal block is passed to, and multiplied with, all off-diagonal blocks in the $I^{th}$ row and $I^{th}$ column. In Phase~3, all remaining blocks $(\hat{I}, J)$ are updated using the product of blocks $(\hat{I}, I)$ and $(I, J)$ from Phase~2.}
   \label{blockfloydphases}
\end{figure}

In our approach, we iteratively update the entire APSP adjacency matrix, $G$, in three phases (Fig.~\ref{blockfloydphases} illustrates in which phase which block of $G$ is updated). In the first phase, we solve sequential Floyd-Warshall restricted to a single sub-block on the diagonal. This is the first diagonal block when iteration $I = 0$. Once we have solved the diagonal block we share the solution with blocks in the same row (and column). This begins Phase~2. Upon receiving the diagonal block, we perform min-plus matrix multiplication to in-place update Phase~2 blocks. In the final phase, all remaining blocks receive two matrices from Phase~2. One matrix is sent from the block in the shared row and the other from the shared column. After computing the min-plus product of the received matrices, we update Phase 3 blocks in-place. The process repeats with successive diagonal blocks. At the conclusion of all 3 phases for the final diagonal block, all matrix entries contain the corresponding shortest path length.


We realize the above algorithm in Spark, starting from the persisted RDD $M$ representing graph $G$. We proceed iteratively over blocks, $G^{(I, I)}$, on the diagonal, which form the critical path of the algorithm. To extract diagonal block $(I,I)$, we use simple \spark{filter} transformation over the keys assigned to each block. We follow with \spark{flatMap} that executes Floyd-Warshall over sub-matrix $G^{(I,I)}$. Upon completion, the transformation yields multiple instances of the updated block with keys $(I, J)$ and $(\hat{I}, I)$, where $0 \leq \hat{I}, J < q$. The keys refer to the blocks in the same row and column where the updated diagonal block is needed for the next phase of computation. Here, similar to $k$NN search, we replicate the data to expose parallelism inherent to~Phase~2.

To begin Phase~2, we again \spark{filter} $G$, this time to obtain sub-blocks with keys $(I, J)$ and $(\hat{I}, I)$. These blocks store values not yet updated for iteration $I$. Thus, we perform \spark{union} transformation to bring together these blocks and the diagonal blocks processed in the earlier Phase. When applying \spark{union} we take special care to ensure that the number of partitions in the resulting RDD is non-increasing. This could negatively impact the subsequent reduction operations, since the exchange of blocks requires shuffle between partitions. Therefore, we use \spark{partitionBy} to ensure all RDDs involved in \spark{union} have same partitioning as $G$. Next, we execute \spark{combineByKey} in order to pair each block from row (column) $I$ with the diagonal block $(I,I)$. To conclude Phase~2, we run \spark{flatMap} that performs min-plus matrix multiplication and in parallel updates the current blocks. The transformation yields additional instances of the updated blocks such that they can be passed to Phase~3. 

To conclude single iteration, we update Phase~3 blocks $G^{(\hat{I}, J)}$, and this involves computing the min-plus product $C = G^{(\hat{I},I)}\cdot G^{(I, J)}$. 
Hence, we use yet another \spark{filter} on $G$ to obtain remaining blocks $G^{(\hat{I}, J)}$, which are in neither row $I$ nor column $I$. These blocks, along with blocks we yield at the end of Phase~2, are brought to a single RDD via \spark{union}, again ensuring appropriate partitioning with \spark{partitionBy}. Next, we use \spark{combineByKey} transformation to bring together three matrices needed to update blocks $(\hat{I}, J)$. This includes the current value of $G^{(\hat{I}, J)}$ and Phase~2 blocks specified above, necessary to compute $C$. Once $C$ is ready, $G^{(\hat{I}, J)}$ is updated by computing element-wise minimum of itself with $C$. 



The three phases are repeated for the next diagonal block. At the conclusion of iteration $I = q - 1$, all pairwise distances represent the approximate manifold distances. To prepare the matrix for subsequent normalization, we square each element of the adjacency matrix to obtain the actual feature matrix $A=G^{\circ 2}$.

In our implementation, the sequential APSP component uses the optimized and efficient SciPy implementation of Floyd-Warshall algorithm. It operates in-place on already allocated memory referenced by NumPy arrays. For min-plus matrix multiplication, we use our own Python implementation. However, our Python code is compiled and optimized to the machine code using the Numba just-in-time compiler, providing excellent performance. Finally, to further improve performance of our min-plus code, we enforce C or Fortran memory layout of matrices as appropriate to maximize cache~usage. 

The overall performance of our APSP hinges on the efficient use of \spark{combineByKey} transformation. While the transformation involves expensive data shuffling, we found that in practice it is more efficient than the available alternatives, like exchanging blocks via \spark{collect/broadcast}~\cite{Vacek2015}, or provisioning in shared secondary storage, e.g., write to HDFS or \spark{SparkFiles.addFile}, and \spark{SparkFiles.get}. This is due to the inefficiency of \spark{collect} or file I/O required for alternative approaches. As the problem size $n$ and block size $b$ increase, these operations become a bottleneck. Spark \spark{collect} requires that the diagonal block and an entire row of blocks be brought to the driver in Phases~1~and~2, respectively. Similarly, the use of secondary storage requires a sequence of file write and read operations to make blocks accessible by appropriate tasks. In our experiments, we found the cost of using heap memory combined with \spark{reduceByKey} for duplication of blocks to be superior to collect to driver or use of a secondary storage. 


The final remark is with respect to RDD lineages (i.e., RDD provenance and execution records). Observe that at each iteration over the diagonal, new RDD describing the entire distance matrix is created. The ancestors of this RDD are all RDDs prescribed by earlier transformations. Consequently, the resulting RDD lineage grows from iteration to iteration, potentially overwhelming the Spark driver. Since the driver is responsible also for scheduling, this hinders performance of the entire algorithm. To address this issue, we \spark{checkpoint} and \spark{persist} $A$ to prune the lineage of RDDs which describe its blocks. Frequency of checkpointing depends on the size of the input data and block size $b$, but in all test cases we found that checkpointing every 10 iterations performs best.

\subsection{Matrix Normalization}\label{ssec:matnorm}

The goal of matrix normalization is to transform $A$ so that it is both row and column centered. In other words, the mean of each row and each column is zero. Centering can be expressed as a linear transformation of the form $A = -\frac{1}{2}H^{T}AH$, where $H = \mathbb{I} - \frac{1}{n}1_{n\times n}$ is centering matrix, and $\mathbb{I}$ is identity matrix. Thus, the update of $A$ requires two matrix-matrix multiplications.

As previously discussed, matrix-matrix multiplication is inefficient in Spark. For this reason, in our implementation we double center $A$ by the direct approach, exploiting the fact that $A$ is symmetric. Specifically, we compute the mean of each column of $A$, and we store all means in the row vector $\mu$, where $\mu_i$ is mean of column~$i$. As means for rows of $A$ are equivalent to $\mu^{T}$, we do not compute them explicitly, and to center $A$ we subtract $\mu$ from each row, and~$\mu^{T}$ from each column of $A$. To finalize the process, we add $\hat{\mu}$ to each entry of $A$, where $\hat{\mu}$ is the global mean computed over all elements of $A$.




To express the above algorithm in Spark, we first transform RDD storing $A$ by \spark{flatMap} that for each block $A^{(I,J)}$ in parallel computes sums of its columns. The transformation yields tuple $(J , \mu_{J})$, and if $I \neq J$ additional tuple $(I, \mu_{I})$ that accounts for processing blocks under the diagonal (recall that $A$ is upper-triangular). Here, $\mu_{J}$ represents vector of $b$ means for the block, and is stored as 1D NumPy array. Next we aggregate the final column sums via \spark{reduceByKey} transformation that applies addition operator over vectors $\mu_{J}$ with shared key~$J$. The vector summation is efficiently implemented in NumPy. At this point, the resulting RDD contains complete information to compute the actual column means as well as the global mean $\hat{\mu}$. Hence, we first sum all elements of the sum vectors into a scalar using single \spark{map} transformation, and then use \spark{reduce} action to obtain the global sum at Spark driver. Additionally, we bring to the driver all column sums by executing \spark{collectAsMap} action. After dividing all the sums by $n$ to obtain the desired means, we \spark{broadcast} them back to Spark executors. Here we note that both \spark{reduce}/\spark{collectAsMap} and \spark{broadcast} actions involve relatively small messages (even for large $n$), and hence this communication pattern is the most direct and efficient to way to exchange computed means between executors. Finally, we conclude the normalization step by transforming RDD that stores $A$ via \spark{map} that in parallel applies previously broadcast means to the local blocks $A^{(I,J)}$. The resulting matrix $A$ is ready for spectral decomposition.

\subsection{Spectral Decomposition}\label{ssec:eigen}

Computing eigenvectors and eigenvalues of feature matrix $A$ is the final step common to all spectral dimensionality reduction methods, as eigenvectors represent the desired low-dimensional representation of $X$. However, larg-scale eigendecomposition is computationally challenging, and the existing parallel eigensolvers are primarily designed for shared memory and multi-/many-core architectures~\cite{Choi1996,Balay2017}. Because to the best of our knowledge currently there are no eigensolvers available for Apache Spark, we develop a new approach derived from the \emph{Power Iteration} method~\cite{Golub2012}. By using simultaneous power iteration over $A$ we are able efficiently extract only the top $d$ eigenpairs at the same time. This is computationally advantageous considering that other approaches require iteratively extracting each eigenpair, and in practical dimensionality reduction applications $d$ is relatively small, e.g.,~$d=3$ when data visualization is the~goal.

\begin{Algorithm}[0.45\textwidth]
    \caption{\textsc{Simultaneous Power Iteration}}
	\begin{algorithmic}[1]
        \REQUIRE $A$, $d$, $l$, $t$
        \ENSURE $(Q_d, \Delta_d)$
        \STATE $V^{1} = \mathbb{I}_{n\times d}$
        \STATE $(Q^{1},R)$ = \textsc{QRdecompose}($V^{1}$)
        \FOR{$i = 2,\ldots, l$} 
            \STATE $V^{i} = A \times Q^{i-1}$
            \STATE $(Q^{i},R)$ = \textsc{QRdecompose}($V^{i}$)
            \IF {$\lVert Q^i - Q^{i-1} \rVert_{2} < t$}
            \STATE \textbf{break};
            \ENDIF 
        \ENDFOR
        \STATE $Q_d = Q^{i}$
        \STATE $\Delta_d = \textrm{diag}(R^{\circ\frac{1}{2}})$
        \RETURN{$(Q_d, \Delta_d)$}
    \end{algorithmic}
    \label{alg:simulpowiter}
\end{Algorithm}

The standard power method (see Alg.~\ref{alg:simulpowiter}) begins with a set of $d$ linearly independent vectors, e.g., the first $d$ standard bases ${\bf e_1}, {\bf e_2},\ldots, {\bf e_d}$. It then considers all vectors simultaneously, and iteratively applies QR decomposition until the process converges to some acceptable level $t$, or the predefined number of iterations $l$. In practice, $l$ and $t$ are selected to achieve accuracy required by given target~application. 

To achieve efficient Spark realization of power method we exploit two facts. First, for large $n$ and reasonable values of~$d$, matrices $V^i$ and $Q^i$ have small memory footprint, and hence incur small sequential processing and communication overheads. Second, QR decomposition has extremely efficient BLAS realizations, available in NumPy. Therefore, it makes sense to assign managing and decomposition of matrix $V^{i}$ to the Spark driver (line 5), while offloading computationally expensive matrix product (line~4) for parallel processing by Spark executors. We note that while MLlib~\cite{Meng2016} provides Spark-based QR decomposition (using tall skinny QR algorithm~\cite{Constantine2011}), it is limited only to MLlib \spark{RowMatrix} RDD that would be inefficient during matrix product stage required by the power method. Consequently, in our implementation, the driver is responsible for keeping track of matrices $Q,R$ and $V$, and checking convergence criteria (line 6). After QR decomposition, the driver broadcasts the entire matrix $Q^i$ to all executors. In this way, we can directly multiply each $b\times b$ block of $A$ with the appropriate block of $Q^{i}$. With this approach we avoid costly replication and shuffle for pairing blocks from each matrix for multiplication. We use \spark{flatMap} to compute block products instead. The transformation yields tuples $((I, 0),C = A^{(I, J)}\cdot Q^{(J, 0)})$ for each block $A^{(I, J)}$ of $A$. For non-diagonal blocks we also yield $((J, 0),C = (A^{(I, J)})^{T}\cdot Q^{(I, 0)})$, to account for upper-triangular storage. Here $C$ is a NumPy 2D array representing one of the block products required for the product $V^{i}$. To obtain the final form of $V^{i}$ we use \spark{reduceByKey} transformation, where key $(I,0)$ is used, with element-wise addition of NumPy 2D arrays. In the final step, we bring $V^{i}$ back to the driver via \spark{collectAsMap}.  

To complete a single iteration, we perform QR factorization of $V^{i}$ on the driver, and test the Frobenius norm of the difference between successive $Q^{i}$ for convergence. At the completion of $l$ iterations or upon convergence, we return the current instance of $Q^{i}$ and $R$, finalizing the spectral decomposition stage, and thus the entire dimensionality reduction process.

\section{Experimental Results}\label{sec:results}


To understand performance characteristics of our approach, we performed a set of experiments on a standalone Apache Spark cluster with 25 nodes and GbE interconnect. Each node in the cluster is equipped with 20-core dualsocket Intel Xeon E5v3 2.3 GHz processor, 64 GB of RAM and a standard SATA hard drive. In all tests, Spark driver was run with 56GB of memory, to account for length of the lineage of RDDs in the APSP loop. For the same reason, we run the driver on a dedicated node separately from Spark executors. We allocated one executor per node using the default configuration for the number of cores, i.e., each executor was using all available cores in a node. All executors were configured to use 56GB out of the available 64GB, with the remaining memory available to the operating system and Python interpreter. We note that we tested different configurations of executor-to-core ratios, across different datasets, without noticeable difference in~performance.

Our entire software is implemented in Apache Spark 2.2 and Python 2.7. Compute intensive linear algebra operations are offloaded via NumPy and SciPy to BLAS library, specifically Intel MKL~2017. Finally, we use Numba~0.35 for optimal min-plus matrix multiplication with just-in-time~compilation.

In all experiments, in the $k$NN stage we used $k=10$ for the neighborhood size. For our test data (see below), such $k$ is large enough to deliver single connected component in the neighborhood graph, and small enough to avoid unnecessary {\em shortcut} edges. In the eigendecomposition stage, we used $t=10^{-9}$ and we allowed a maximum of $l=100$ iterations to achieve convergence. We note that in all experiments, many fewer iterations were required to see convergence (usually around 20-50 iterations).


\subsection{Test Data and Method Correctness}

The Swiss Roll benchmark is a classic test for evaluating manifold learning algorithms. Swiss Roll data is generated from some input 2D data, which is embedded into 3D using a non-linear function. In our tests, we used the \emph{Euler Isometric Swiss Roll}~\cite{Schoeneman2017}. To evaluate the correctness and demonstrate the scalability of our implementation, we created three test datasets. These are samples from the Swiss Roll of size $n=50000$, $n=75000$, and $n=100000$, which we refer to as Swiss50, Swiss75, and Swiss100.

\begin{figure*}[htbp!]
        \subfigure[2D data prior to embedding in 3D.]{%
        \includegraphics[width=.325\linewidth,trim={0cm 0cm 0cm 0cm},clip]{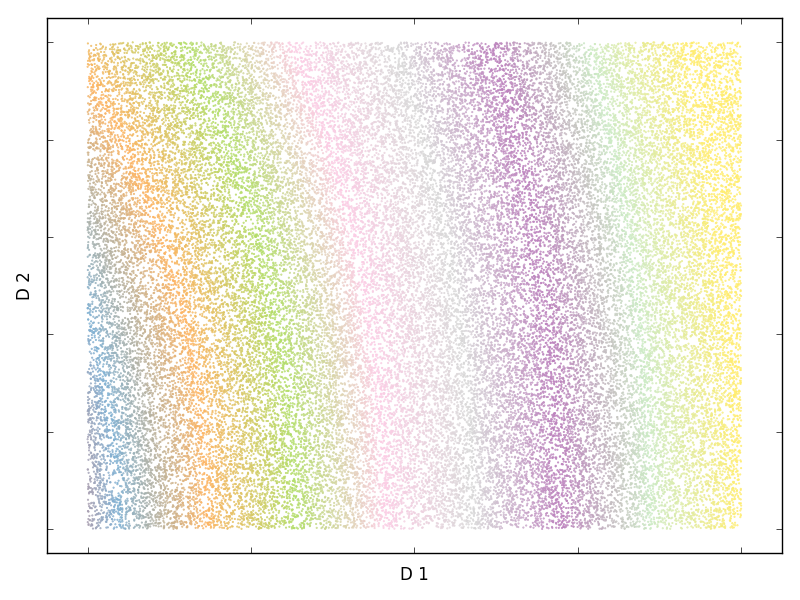}
        \label{fig:swiss50_gt}\hfill
        }
		\subfigure[Data embedded in 3D.]{%
        \includegraphics[width=.325\linewidth,trim={0.35cm 0.35cm 0.35cm 0.35cm},clip]{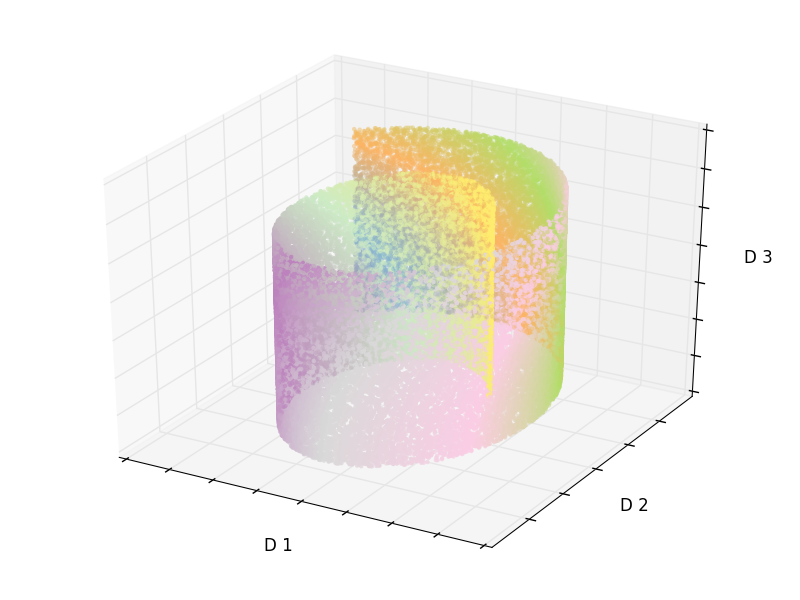}
        \label{fig:swiss50_3d}\hfill
        }
        \subfigure[2D mapping learned by our method, $k=10$.]{%
        \includegraphics[width=.325\linewidth,trim={0cm 0cm 0cm 0cm},clip]{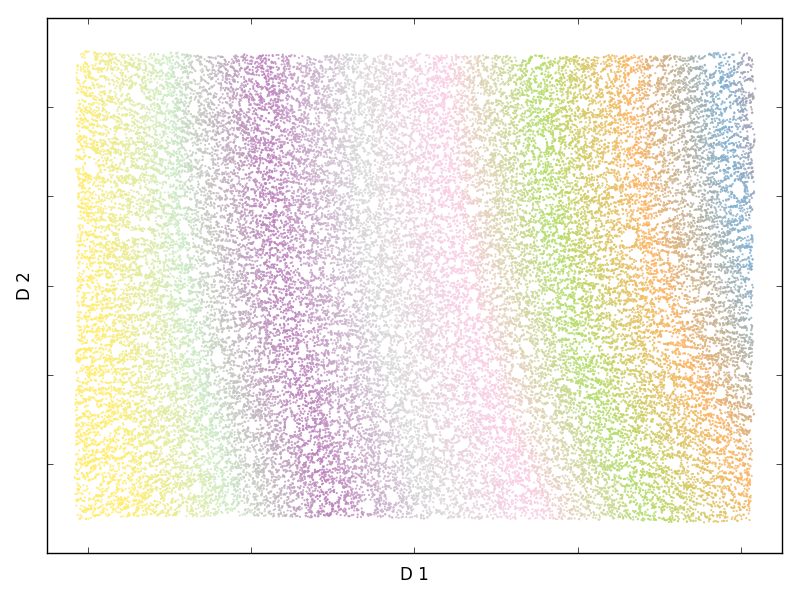}
        \label{fig:swiss50_iso}\hfill
        }
    \caption{To demonstrate the correctness of our Spark Isomap method we sample 50,000 points from the Euler Isometric Swiss Roll, and then perform dimensionality reduction. (Please view in color).
    }
    \label{IsoSpark_swiss50_correct}
\end{figure*}

Figure \ref{IsoSpark_swiss50_correct} shows the initial data, high-dimensional embedded data, and the resulting dimensionality reduction delivered by our method for the Swiss50 dataset. A quick visual inspection reveals that Fig.~\ref{fig:swiss50_iso} appears to be a faithful representation of the original data in Fig.~\ref{fig:swiss50_gt}. To actually quantify how well our method reconstructs the original data, we use \emph{Procrustes error}, which captures the similarity of two datasets $X$ and $Y$ by determining the optimal affine transformation of $X$ that best conforms it with~$Y$~\cite{Dryden1998}. In our case, Procrustes error of our result compared to the original 2D coordinates is $0.000026741$, confirming high accuracy of the reconstruction.

To test our platform on the actual high-dimensional data, we used benchmarks derived from the EMNIST~\cite{Cohen2017}, which is frequently used in machine learning research. This dataset provides $28\times 28$ images of over 200,000 handwritten digits, and hence each data point (image of a digit) is a 784-dimensional vector. For testing purposes, we randomly sampled two sets from EMNIST. They contain $n=50000$ and $n=125000$ points and we refer to them as EMNIST50 and EMNIST125, respectively. 

\begin{figure}[t]
   \centering
   \subfigure[Points cluster by digit which they represent.]{%
      \includegraphics[width=0.75\linewidth]{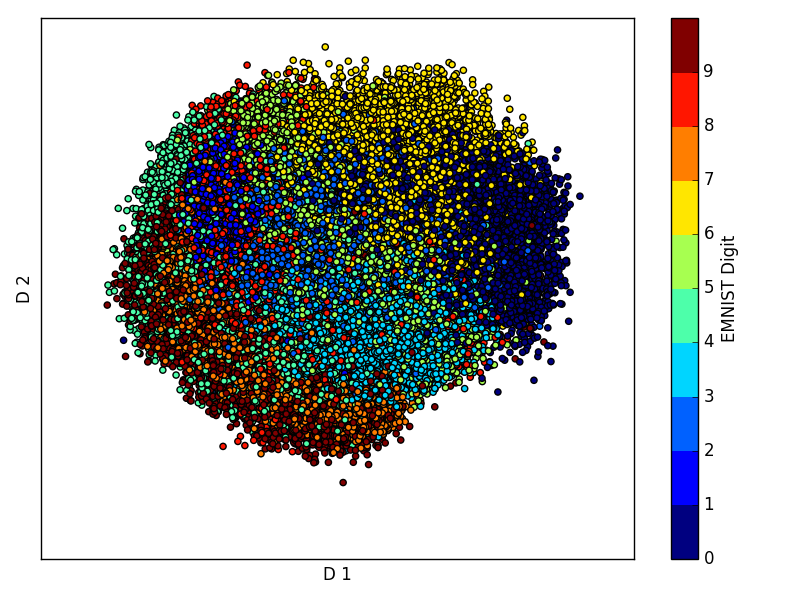}
   \label{fig:emn50_1}\hfill
   }
   \subfigure[Sample of original images shown to highlight features captured by reduced dimensions D1 and D2.]{%
   \includegraphics[width=0.75\linewidth]{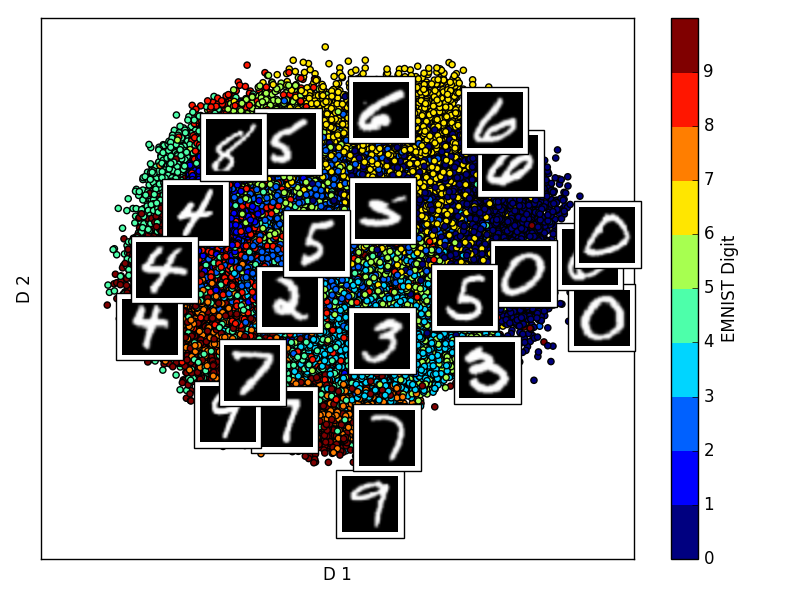}
   \label{fig:emn50_2}\hfill
   }
   \caption{To demonstrate our method on high-dimensional data we sample 50,000 points from the EMNIST dataset $(D=784)$, and then perform dimensionality reduction $(d=2, k=10)$. Original image shown for selected points. (Please view in color).}
   \label{emnist50k2dimg}
\end{figure}

Figure \ref{emnist50k2dimg} shows 2D mapping of EMNIST50 obtained using our Isomap method. From the figure we can see that clusters of digits that look similar appear close together, e.g., `5'~and~`6'. At the same time, clusters of distinct digits that blend together also have these similarities. In Fig.~\ref{fig:emn50_2} sample images of digits are included to accentuate features captured by axes D1 and D2. The axis D2 describes the angle of slant for the handwritten digit. We observe the change in angle from top to bottom of the cluster. The primary axis, D1, accounts for curved or straight segments in the digit. For example, on the left we see `4's, which are made up entirely of straight line segments. In contrast, zeros on the right contain no straight segments. To the best of our knowledge, this is the first time such result is reported, and it confirms practical value of large-scale Isomap in processing noise image data.

\subsection{Scalability Tests}

To analyze how our implementation scales with the problem size, expressed by $n$, and the compute resources used, we processed all datasets on a varying number of nodes, recording the wall time. The results of this experiment are summarized in Tables~\ref{exectimetable}-\ref{efftable}. Here we report relative speedup, computed as a fraction $S_p = \frac{T_{min}}{T_p}$, where $T_p$ is time taken to process data instance on $p$ nodes, and $T_{min}$ is the time taken by the same instance to process on the smallest feasible number of nodes. We note that even the smallest datasets we consider in our tests are impossible to process using single node, and hence $T_{min}$ is not $T_{1}$. Similarly, we compute relative efficiency as $E_p = \frac{S_p}{p} \cdot \arg{T_{min}}$. 

The experimental results show that our approach exhibits strong scaling within the entire range of data and cluster configuration. For smaller datasets ($n=50000$), the method scales linearly up to 12 nodes, and maintains very good efficiency up to 24 nodes. The scaling is very similar for both Swiss50 and EMNIST50, which is expected taking into account that only $k$NN stage depends (linearly) on $D$, and it constitutes only a small fraction of the entire runtime. At this point we should note that because we express all Isomap stages as linear equations or matrix multiplications, the performance of our method is not at all impacted by data distribution or density (e.g., connectivity of graph $G$). When the size of the input data increases, the method seems to exhibit super-linear performance, however, this result should be taken with care, considering that we compute relative and not actual speedup. The method maintains also weak scalability. Specifically, for the fixed ratio $\frac{n}{p}$, the execution time scales roughly as $\left( \frac{n}{p} \right)^3$, which reflects the dominating cost $\mathcal{O}(n^{3})$ of the APSP stage.

\begin{table}[H]
\caption{Execution time in minutes}
\label{exectimetable}
\centering
{\footnotesize
\begin{tabularx}{\columnwidth}{lXXXXXXX}
\toprule
& \multicolumn{7}{c}{Compute Nodes}\\
Name & 2 & 4 & 8 & 12 & 16 & 20 & 24\\
\midrule
EMNIST50 & 294.92 & 117.90 & 62.41 & 48.47 & 41.44 & 35.82 & 32.92\\
Swiss50 & 261.43 & 98.88 & 53.58 & 39.41 & 33.26 & 29.47 & 26.49\\
Swiss75 & -- & 958.85 & 345.47 & 105.28 & 84.05 & 76.66 & 63.25\\
Swiss100 & -- & -- & 1122.89 & 465.30 & 275.22 & 225.89 & 157.97\\
EMNIST125 & -- & -- & -- & 985.94 & 662.92 & 599.28 & 448.42\\
\bottomrule
\end{tabularx}
{\footnotesize -- means that dataset was impossible to process on given resources.}
}
\end{table}%
\begin{table}[H]
\caption{Relative speedup}
\label{spduptable}
\centering
{\footnotesize
\begin{tabularx}{\columnwidth}{lXXXXXXX}
\toprule
& \multicolumn{7}{c}{Compute Nodes}\\
Name & 2 & 4 & 8 & 12 & 16 & 20 & 24\\
\midrule
EMNIST50 & 1 & 2.50 & 4.73 & 6.09 & 7.11 & 8.23 & 8.96\\
Swiss50 & 1 & 2.64 & 4.88 & 6.63 & 7.86 & 8.87 & 9.87\\
Swiss75 & -- & 1 & 2.78 & 9.11 & 11.41 & 12.51 & 15.16\\
Swiss100 & -- & -- & 1 & 2.41 & 4.08 & 4.97 & 7.11\\
EMNIST125 & -- & -- & -- & 1 & 1.49 & 1.65 & 2.20\\
\bottomrule
\end{tabularx}
}
\end{table}%
\begin{table}[H]
\caption{Relative efficiency}
\label{efftable}
\centering
{\footnotesize
\begin{tabularx}{\columnwidth}{lXXXXXXX}
\toprule
& \multicolumn{7}{c}{Compute Nodes}\\
Name & 2 & 4 & 8 & 12 & 16 & 20 & 24\\
\midrule
EMNIST50 & 1 & 1.25 & 1.18 & 1.02 & 0.89 & 0.82 & 0.75\\
Swiss50 & 1 & 1.32 & 1.22 & 1.11 & 0.98 & 0.89 & 0.82\\
Swiss75 & -- & 1 & 1.39 & 3.04 & 2.85 & 2.50 & 2.53\\
Swiss100 & -- & -- & 1 & 1.61 & 2.04 & 1.99 & 2.37\\
EMNIST125 & -- & -- & -- & 1 & 1.12 & 0.99 & 1.10\\
\bottomrule
\end{tabularx}
}
\end{table}

\begin{figure}[t]
   \centering
   \includegraphics[width=0.75\linewidth]{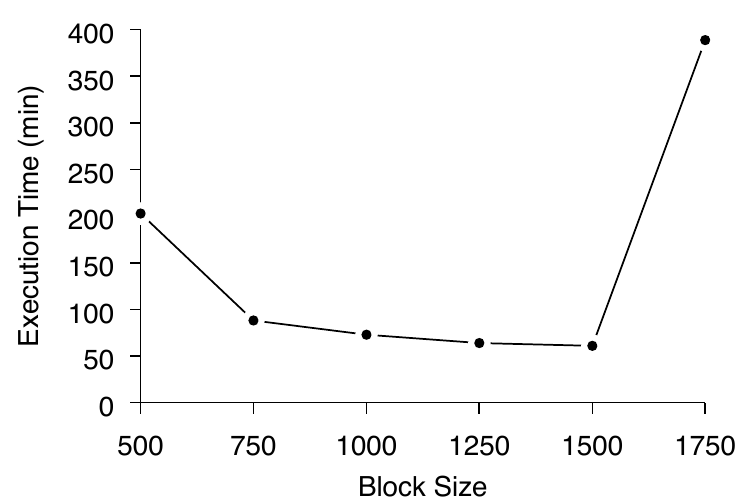}
   \caption{To identify the effect of block size on overall execution time, we run our method on Swiss75 using 24 compute nodes with varying $b$.}
   \label{bvstime}
\end{figure}

The performance of our method directly depends on the logical block size $b$ (see Sec.~\ref{ssec:knn}). Recall that in $k$NN search, blocks of $b$ points are paired to compute one block of the pairwise distance matrix. The execution time of computing a single block scales as $\Theta$($b^{2}$). When computing APSP, we require sequential Floyd-Warshall for each block along the diagonal, and matrix-matrix min-plus multiplication for others. Both cases have complexity $\Theta(b^{3})$. Moreover, the APSP solver proceeds iteratively over the diagonal~of~$A$. Since each block on the diagonal must be solved in sequence, the critical path has length $q$. All these factors must be taken into consideration when selecting $b$. In our experiments, we found that $b$ in the range $1000\leq b \leq 2500$ gives ideal performance when $n \leq 100,000$. This is because the length of the critical path is not overwhelming for the Spark driver, and at the same time, the block size is small enough to leverage cache memory when executing BLAS routines for matrix products. As the problem size grows, it is advantageous to increase block size to keep control of the critical path. This however may increase the time taken to process individual block, and may lead to resource underutilization as there are fewer blocks to distribute for processing between executors. Figure~\ref{bvstime} shows impact of block size $b$ when processing Swiss75 on 24 nodes (we observe similar pattern for other datasets). The sweet spot is for $b=1500$ and both undersizing and oversizing $b$ degrades performance. Currently, we do not have model that would let us select $b$ automatically, however, the intuition we provide above is sufficient for the majority of practical applications.

\section{Related Work}\label{sec:relwork}

As previously mentioned, existing implementations of Isomap and its variants do not scale to data as large as presented here. Talkwalkar et al. successfully learned manifold for 18 million points. However, their analysis includes approximation method for APSP, and Nystrom method for approximate spectral decomposition. To date, the largest study for exact solutions for Isomap includes $n=32768$ points, and has been reported in~\cite{Samudrala2015}. The method is tailored for HPC clusters with MPI, and hence its applicability is somewhat constrained.

To partially mitigate the complexity of APSP, the authors of Isomap have proposed Landmark-Isomap (L-Isomap)~\cite{deSilva2003}. The idea behind L-Isomap is that $m$ randomly selected {\em landmark points} may be selected to learn an underlying manifold. Remaining points are placed on the manifold via triangulation using their distance to landmarks. This approach greatly reduces complexity by targeting APSP, but introduces new sources of error in i) landmark selection method, ii) selecting the number of landmarks, and iii) adding the complete dataset by triangulation. We previously extended the works of de Silva and Tenenbaum~\cite{deSilva2003,Tenenbaum2000} and proposed Streaming-Isomap~\cite{Schoeneman2017}. The streaming approach requires learning a faithful representation of the underlying manifold for some initial batch of points, and then quickly maps new points arriving on a high-volume, high-throughput stream. This approach is orthogonal to the one we present here, and in fact both methods could be combined in case when the initial batch is large.

\section{Conclusion}\label{sec:end}

High-dimensional big data, arising in many practical applications, presents a challenge for non-linear spectral dimensionality reduction methods. This is due to the computationally intensive construction of the feature matrix and it's spectral decomposition. In this paper, we proposed a scalable distributed memory approach for manifold learning using Isomap, addressing key computational bottlenecks of the method. Through experiments we demonstrated that our resulting Apache Spark-based implementation is scalable and efficient. The method can solve exact Isomap on high-dimensional datasets, which are an order of magnitude larger than what can be processed with the existing methods. We note that individual components, like $k$NN, APSP and eigendecomposition solvers, can be used as a standalone routines. Finally, since other non-linear spectral decomposition methods, like e.g., LLE~\cite{Roweis2000}, share the same computational backbone, with a minimal effort our software could be extended to cover these methods as well. The source code of our platform is open and available from: \url{https://gitlab.com/SCoRe-Group/IsomapSpark}.


\section*{Acknowledgment}
The authors would like to acknowledge support provided by the Center for Computational Research at the University at Buffalo. This work has been supported in part by the Department of Veterans Affairs under grant 7D0084.



%




\bibliographystyle{IEEEtran}
\bibliography{sample-bibliography.bib}

\begin{thebibliography}{10}
\providecommand{\url}[1]{#1}
\csname url@samestyle\endcsname
\providecommand{\newblock}{\relax}
\providecommand{\bibinfo}[2]{#2}
\providecommand{\BIBentrySTDinterwordspacing}{\spaceskip=0pt\relax}
\providecommand{\BIBentryALTinterwordstretchfactor}{4}
\providecommand{\BIBentryALTinterwordspacing}{\spaceskip=\fontdimen2\font plus
\BIBentryALTinterwordstretchfactor\fontdimen3\font minus
  \fontdimen4\font\relax}
\providecommand{\BIBforeignlanguage}[2]{{%
\expandafter\ifx\csname l@#1\endcsname\relax
\typeout{** WARNING: IEEEtran.bst: No hyphenation pattern has been}%
\typeout{** loaded for the language `#1'. Using the pattern for}%
\typeout{** the default language instead.}%
\else
\language=\csname l@#1\endcsname
\fi
#2}}
\providecommand{\BIBdecl}{\relax}
\BIBdecl

\bibitem{Lee2007}
J.~Lee and M.~Verleysen, \emph{Nonlinear Dimensionality Reduction}.\hskip 1em
  plus 0.5em minus 0.4em\relax Springer Verlag, 2007.

\bibitem{Tenenbaum2000}
J.~Tenenbaum, V.~de~Silva, and J.~Langford, ``A global geometric framework for
  nonlinear dimensionality reduction,'' \emph{Science}, vol. 290, no. 5500, p.
  2319, 2000.

\bibitem{Bernstein2000}
M.~Bernstein, V.~de~Silva, J.~Langford, and J.~Tenenbaum, ``Graph
  approximations to geodesics on embedded manifolds,'' 2000.

\bibitem{Turnbaugh2007}
P.~Turnbaugh, R.~Ley, M.~Hamady, C.~Fraser-Liggett, R.~Knight, and J.~Gordon,
  ``The human microbiome project,'' \emph{Nature}, vol. 449, no. 7164, p. 804,
  2007.

\bibitem{Marchand2009}
R.~Marchand, N.~Beagley, and T.~Ackerman, ``Evaluation of hydrometeor
  occurrence profiles in the multiscale modeling framework climate model using
  atmospheric classification,'' \emph{Journal of Climate}, vol.~22, no.~17, pp.
  4557--4573, 2009.

\bibitem{Thirion2004}
B.~Thirion and O.~Faugeras, ``Nonlinear dimension reduction of {fMRI} data: the
  {Laplacian} embedding approach,'' in \emph{IEEE International Symposium on
  Biomedical Imaging: Nano to Macro}, 2004, pp. 372--375.

\bibitem{Li2012}
W.~Li, S.~Prasad, J.~Fowler, and L.~Bruce, ``Locality-preserving dimensionality
  reduction and classification for hyperspectral image analysis,'' \emph{IEEE
  Transactions on Geoscience and Remote Sensing}, vol.~50, no.~4, pp.
  1185--1198, 2012.

\bibitem{deSilva2003}
V.~de~Silva and J.~Tenenbaum, ``Global versus local methods in nonlinear
  dimensionality reduction,'' in \emph{Advances in Neural Information
  Processing Systems}, 2003, pp. 721--728.

\bibitem{Talwalkar2008}
A.~Talwalkar, S.~Kumar, and H.~Rowley, ``Large-scale manifold learning,'' in
  \emph{IEEE Conference on Computer Vision and Pattern Recognition}, 2008, pp.
  1--8.

\bibitem{Samudrala2015}
S.~Samudrala, J.~Zola, S.~Aluru, and B.~Ganapathysubramanian, ``Parallel
  framework for dimensionality reduction of large-scale datasets,''
  \emph{Scientific Programming}, vol. 2015, p.~1, 2015.

\bibitem{Bentley1975}
J.~Bentley, ``Multidimensional binary search trees used for associative
  searching,'' \emph{Communications of the ACM}, vol.~18, no.~9, pp. 509--517,
  1975.

\bibitem{Kim2016}
W.~Kim, Y.~Kim, and K.~Shim, ``Parallel computation of k-nearest neighbor joins
  using {MapReduce},'' in \emph{IEEE International Conference on Big Data},
  2016, pp. 696--705.

\bibitem{Weber1998}
R.~Weber, H.~Schek, and S.~Blott, ``A quantitative analysis and performance
  study for similarity-search methods in high-dimensional spaces,'' in
  \emph{International Conference on Very Large Data Bases}, 1998, pp. 194--205.

\bibitem{Meng2016}
X.~Meng, J.~Bradley, B.~Yavuz, E.~Sparks, S.~Venkataraman, D.~Liu, J.~Freeman,
  D.~Tsai, M.~Amde, S.~Owen \emph{et~al.}, ``{MLlib}: Machine learning in
  apache spark,'' \emph{The Journal of Machine Learning Research}, vol.~17,
  no.~1, pp. 1235--1241, 2016.

\bibitem{Cormen2009}
T.~Cormen, C.~Leiserson, R.~Rivest, and C.~Stein, \emph{Introduction to
  algorithms}.\hskip 1em plus 0.5em minus 0.4em\relax MIT press, 2009.

\bibitem{Kepner2011}
J.~Kepner and J.~Gilbert, \emph{Graph Algorithms in the Language of Linear
  Algebra}.\hskip 1em plus 0.5em minus 0.4em\relax Society for Industrial and
  Applied Mathematics, 2011.

\bibitem{Bosagh2016}
R.~Zadeh, X.~Meng, A.~Ulanov, B.~Yavuz, L.~Pu, S.~Venkataraman, E.~Sparks,
  A.~Staple, and M.~Zaharia, ``Matrix computations and optimization in {Apache
  Spark},'' in \emph{ACM SIGKDD International Conference on Knowledge Discovery
  and Data Mining}, 2016, pp. 31--38.

\bibitem{Solomonik2013}
E.~Solomonik, A.~Buluc, and J.~Demmel, ``Minimizing communication in all-pairs
  shortest paths,'' in \emph{IEEE International Symposium on Parallel and
  Distributed Processing}, 2013, pp. 548--559.

\bibitem{Venkataraman2003}
G.~Venkataraman, S.~Sahni, and S.~Mukhopadhyaya, ``A blocked all-pairs
  shortest-paths algorithm,'' \emph{Journal of Experimental Algorithmics},
  vol.~8, 2003.

\bibitem{Vacek2015}
T.~Vacek, ``Flyby improved dense matrix multiplication,'' 2015.

\bibitem{Choi1996}
J.~Choi, J.~Demmel, I.~Dhillon, J.~Dongarra, S.~Ostrouchov, A.~Petitet,
  K.~Stanley, D.~Walker, and R.~Whaley, ``{ScaLAPACK}: A portable linear
  algebra library for distributed memory computers—design issues and
  performance,'' \emph{Computer Physics Communications}, vol.~97, no. 1-2, pp.
  1--15, 1996.

\bibitem{Balay2017}
S.~Balay, S.~Abhyankar, M.~Adams, J.~Brown, P.~Brune, K.~Buschelman, L.~Dalcin,
  V.~Eijkhout, W.~Gropp, D.~Kaushik \emph{et~al.}, ``{PETSc} users manual
  revision 3.8,'' Argonne National Laboratory (ANL), Tech. Rep., 2017.

\bibitem{Golub2012}
G.~Golub and C.~V. Loan, \emph{Matrix Computations}.\hskip 1em plus 0.5em minus
  0.4em\relax JHU Press, 2012.

\bibitem{Constantine2011}
P.~Constantine and D.~Gleich, ``Tall and skinny {QR} factorizations in
  {MapReduce} architectures,'' in \emph{International Workshop on MapReduce and
  Its Applications}, 2011, pp. 43--50.

\bibitem{Schoeneman2017}
F.~Schoeneman, S.~Mahapatra, V.~Chandola, N.~Napp, and J.~Zola, ``Error metrics
  for learning reliable manifolds from streaming data,'' in \emph{SIAM
  International Conference on Data Mining}, 2017, pp. 750--758.

\bibitem{Dryden1998}
I.~Dryden and K.~Mardia, \emph{Statistical Shape Analysis}.\hskip 1em plus
  0.5em minus 0.4em\relax John Wiley \& Sons, 1998.

\bibitem{Cohen2017}
G.~Cohen, S.~Afshar, J.~Tapson, and A.~van Schaik, ``{{EMNIST}: an extension of
  {MNIST} to handwritten letters},'' \emph{ArXiv e-prints}, 2017.

\bibitem{Roweis2000}
S.~Roweis and L.~Saul, ``Nonlinear dimensionality reduction by {Locally Linear
  Embedding},'' \emph{Science}, vol. 290, no. 5500, pp. 2323--2326, 2000.

\end{thebibliography}

\end{document}